# Ferromagnetism in Mn Substituted Zirconia: A Density-functional Theory Study

Xingtao Jia[1], Wei Yang[2], Minghui Qin[3], Xinglai Zhang[2], Mingai Sun[2], Jianping Li[4]

[1] College of Chemistry & Chemical Engineering, China University of Petroleum, Dongying, 257061, People's Republic of China

[2] College of Physics Science and Technology, China University of Petroleum, Dongying, 257061, People's Republic of China

[3] National Laboratory of Solid State Microstructures and Department of Physics, Nanjing University, Nanjing 210093, People's Republic of China

[4] deparment of precision instruments, Shandong College of Information Techonology, Weifang, 261041, People's Republic of China

E-mail: yangwei_upc@yahoo.com.cn

**Abstract.** We study the electronic structure and magnetism of 25% Mn substituted cubic Zirconia ($ZrO_2$) with several homogeneous and heterogeneous doping profiles using density-functional theory calculations. We find that all doping profiles show half-metallic ferromagnetism (HMF), and delta-doping is most energy favorable while homogeneous doping has largest ferromagnetic stabilization energy. Using crystal field theory, we discuss the formation scheme of HMF. Finally, we speculate the potential spintronics applications for Mn doped $ZrO_2$, especially as spin direction controllment.

## 1. Introduction

Silicon-based Spintronics has attracted much attention recently, especially the realization of remarkable spin injection and transport in silicon [1-5], which realizes the seamless fusion of semiconductor and spintronics technology and inspires the forthcoming of a new semiconductor-spintronics paradigm. The most intensively studied spintronics materials can classify into three types: magnetic metals and alloys, Heusler alloys, and magnetic semiconductors (MS) [6]. Among them, magnetic semiconductor with half-metallic ferromagnetism (HMF) is regarded as the most promising candidate for silicon-based spintronics. Recently, the researches about wide band-gap insulators ($E_g > 4$ eV) have attracted much attention for the conceivable high temperature application in silicon-based spintronics [7-14], indicates that magnetic insulators have joined in the spintronics club.

    For magnetic semiconductors and magnetic insulators, the most obstacles to spintronics application would be how to realize credible magnetism. Although many synthetic and theoretical studies show that a large group of semiconductors and insulators can realize room or higher temperature magnetism by doping with magnetic impurities; unfortunately the magnetism is delicate to the synthesis conditions and hard to reproduce. Extensive studies show that magnetic atoms are energetic favorable to agglomerate into clusters, which would degenerate the magnetism and make the materials impracticable [15-18]. So, a stable doping profile is important for spintronics application.

Delta-doping is a kind of heterogeneous doping profile with one or less substituted atoms layer at a specific face. Compared with homogeneous or other heterogeneous doping profiles, delta-doping is easy to realize by molecular-beam epitaxy (MBE) synthesis and widely used in current semiconductor industry [19,20], especially for polarized materials. Compared with non-polarized materials, polarized materials would be stabilized by the existence of polarization stabilization energy (energy difference between positive and negative polarity faces). Moreover, delta-doping always shows enhanced properties [21-25]. So, we speculate that delta-doping would be more useful (in spintronics application) than other doping profiles.

Wide-band dielectric materials, such as Hafnia ($HfO_2$) and Zirconia ($ZrO_2$), have been integrated into the 45 nm paradigm silicon-based integrated circuit as the key insulating gate materials in complementary metal-oxide-semiconductor field effect transistors [26-28]. Recent studies show that $HfO_2$ and $ZrO_2$ demonstrate HMF and are supposed to be potential spintronics applications when doped with magnetic atoms [7-13]. Besides, an obvious advantage for half-metallic $HfO_2$ and $ZrO_2$ based spintronics materials would be good compatibility with both group IV semiconductors and metals.

In the study, on the basis of *state-of-the-art ab initio* electronic structure calculations, we investigate the electronic and magnetic properties of Mn substituted $ZrO_2$ with several homogeneous and heterogeneous doping profiles. Mainly, there are three crystal structures for $ZrO_2$. At ambient temperature, $ZrO_2$ show monoclinic structure and transform to tetragonal structure (> 1400 K) and then to cubic fluorite structure (> 2600 K) as temperature increase. Cubic fluorite $ZrO_2$ show higher dielectric constant than monoclinic and tetragonal phase, and draw more attention in insulating gate use. Theoretically, cubic $ZrO_2$ show distinct polarized face and would be easy epitaxial growth on polarized substrates. Experimentally, cubic $ZrO_2$ thin films have been epitaxial growth on group IV semiconductors and alloys [29,30]; by doping with transition metal atoms, cubic phase would be stabilized to room temperature [31]. Ostanin and coworkers [7] studied the electronic structure and magnetism of cubic $ZrO_2$ doped with a series of magnetic atoms, and concluded the possibility of spintronics application. This is a very creative study. But unfortunately, ideal homogeneously doping is hard to realized experimentally. In the study, we find that delta-doping shows robust HMF, and is more thermodynamic stable than homogeneous and other heterogeneous doping profiles. Finally, based on the simply crystal filed theory (CFT), we investigate the magnetism and stabilization mechanism in delta-doping profile.

## 2. Computational details

We employed plane-wave pseudopotential density functional theory (DFT) calculation, where the spin-dependent generalized gradient approximation (GGA) is used for the exchange and correlation effects [32]. In the total energy calculations, Ultrasoft pseudopotentials proposed by Vanderbilt were used to describe the ionic potentials of TM atoms [33], and the exchange-correlation functional parameterized by Perdew, Burke, and Ernzerhof (PBE) [34]. To simplify the calculation, a 1x1x2 cubic $ZrO_2$ supercell with two substituted Mn atoms (as shown in figure 1) was used in the calculation. A plane-wave basis energy cutoff of 450eV, and Monkhorst-Pack grid of 6×6×3 over the Brillouin zone were used to ensure the convergence in the calculations. All compounds were optimized with relaxation of both lattice parameters and atomic positions. The total energy was converged to $1.0 \times 10^{-5}$ eV/atom while the Hellman-Feynman force was $3 \times 10^{-2}$ eV/Å in the optimized structure.

## 3. Results and discussion

The calculated total magnetic moment, total energy (with respect to N112), and ferromagnetic stabilization energy in full relaxed Mn substituted $ZrO_2$ with different doping profiles are given in table I. Apparently most doping profiles show large ferromagnetic stabilization energy except N112, indicating robust ferromagnetism in these doping profiles. Generally, the doping profiles show little effect on the total magnetic moment, while great on the total energies of the studied doping profiles. The total energy shows close relation with the distance between Mn-Mn pairs: the nearer the distance, the lower the energy. Only a slight deviation exists in N011 and N110 configuration (also for N002 and N112), where the Mn-Mn pair distances are close but the total energies are different. This may be attributed to more neighbor Mn atoms (four) around one Mn in N110 than in N011 (two); (for N002 and N112, there are six and four neighbor Mn atoms respectively). Unlike total energies, the ferromagnetic stabilization energy is more related with the number of neighbor Mn atoms in four doping profiles. In a word, the more homogeneous, the larger the ferromagnetic stabilization energy; the closer the Mn-Mn distance, the lower the total energy.

The spin-resolved total Density of states (DOS) of full relaxed Mn doped $ZrO_2$ with different doping profiles are shown in figure 2. Apparently all four studied profiles show pronounced HMF, this is consistent with the integral value of total magnetic moment (Table 1) and the study by Ostanin and cooperators [7]. Under high doping concentration, the energy levels of Mn impurity would span into bands, which would hybridize with bulk $ZrO_2$ further. But the minority-spin DOS in the band-gap of bulk $ZrO_2$ are mainly from the Mn impurity. The impurity DOS around the $E_F$ show clear peaks for all four doping profiles, but different in details such as the position of peaks, the numbers, and the fade line and so on. For the minority-spin channel, there are two clusters of impurity DOS peaks above the top of valence band separated by the pseudogap for all doping profiles, which can be associated with the $e_g$-$t_{2g}$ splitting expected from crystal field model, also show obvious difference in details.

To illustrate the formation mechanism of HMF in Mn doped $ZrO_2$, a simple crystal field model is schemed as shown in figure 3. The principal mechanism leading to the appearance of the half-metallic gap is the hybridization of the Mn $d$ orbitals with O $p$ orbitals. Here, the $p$-$d$ hybridization can be dictated by the cubic field, where each Mn is surrounded by eight O atoms. The cubic filed can be divided into two tetrahedral fields, where the $d$ orbitals would split into triple-degenerated $t_{2g}$ orbitals (consisting of $d_x$, $d_y$ and $d_z$ orbitals) and double-degenerated $e_g$ orbitals (consisting of $d_{x^2-y^2}$ and $d_{z^2}$ orbitals). For spatial symmetry, only Mn $t_{2g}$ orbitals can hybridize with the neighbor O $p$ orbitals, while Mn $e_g$ orbitals remain unhybridized. Actually, the Mn $d$ orbitals are about 4.6 eV higher than the O $p$ orbitals. So, when Mn $d$ hybridizes with O $p$ orbitals, the lower hybridized $t_{2g}$ orbitals would be dominated by O $p$ orbitals while the higher by Mn $d$ orbitals. This would be the reason than the unoccupied minority $e_g$ and $t_{2g}$ states are mainly from Mn $d$ states in Fiture 2. Moreover, any distortion in crystal field would affect the band hybridization in Mn doped $ZrO_2$.

The doping profiles show obviously impact on the DOS of Mn doped $ZrO_2$ as shown in figure 2. This originates from the Jahn-Teller distortion introduced by heterogeneous doping. For minority-spin channel, The minority-spin $e_g$ and $t_{2g}$ states lying in the forbidden gap of bulk $ZrO_2$ show obvious difference in details among the four doping profiles. There are one intensive and one degenerated $e_g$ peaks (also for $t_{2g}$ states) for N002, while two intensive $e_g$ peaks and two fingerlike $t_{2g}$ peaks, indicates that the $e_g$ states of N112 are more localized than N002. The $e_g$ and $t_{2g}$ states of N011 show similarity with N002 except a broadened $e_g$ peaks. For N110, there shows a broadened two peak $e_g$ states and a separated two peak $t_{2g}$ states (also shown in figure 4). Interestingly, the $e_g$ and $t_{2g}$ states of N002 are melt into one impurity states, while N112 and N110 are separate, indicates that the $e_g$-$t_{2g}$ splitting is enhanced in N112 and N110. For the majority-spin channel, the peaks around the $E_F$ can be associated with $t_{2g}$ orbitals. Except the number, position, and intensity of the fingerprint peaks, the remarkable difference lies in the fadeline of $t_{2g}$ states. The fadeline of $t_{2g}$ states above the $E_F$ is 0.85, 0.97, 0.93, and 1.15 eV for N112, N002, N011 and N110 respectively. According to Hordequin et al [35] and Qian et al [25], when the $E_F$ lying near the top of the valence band of the minority-spin channel and the majority-spin DOS fadeline extend to a larger value, the spin-flip transition temperature would reach to as high as Curie temperature ($T_C$). Ostanin and coworkers [7] calculated the relation of $T_C$ and Mn concentration in $ZrO_2$ using the KKR-CPA method; they found the $T_C$ would reach to

570 K at 25% Mn concentration. Interestingly, our calculations show that the fadeline of majority-spin DOS around the $E_F$ would extend to 1.15 eV under N110 doping profile, which is rather higher than corresponding homogenous doping N002 (0.97 eV), indicates an even higher $T_C$ would be under the N110 profile. One thing should be mention, the $T_C$ here shows inconsistent with which expected from the mean field theory, where the $T_C$ relate with $E_{FA}$. So, the practical experiments should conduct to check which is right.

One big problem for doping would be how to realize, especially for homogeneous doping. Compared with homogenous doping, delta-doping would be easy to realize for the development of modern MBE technology. In the study, N110 profile is a kind of delta-doping at $ZrO_2$ (001). As discussed above, delta-doping is energy favorable and HMF robustness than homogeneous doping (and other kinds of heterogeneous doping profiles). That is, delta-doping may be a more practical method for magnetic semiconductors.

Figure 4 show the total, partial and element resolved DOS of the delta-doping (N110) profile. As discussed above, we can see that *s* orbital contribute little to the total DOS, while the bonding $t_{2g}$ and $e_g$ states for both spin channels are from O *p* and Mn *d* orbitals, where the O *p* orbitals are dominating. Apparently, the unoccupied minority $e_g$ and $t_{2g}$ states are mainly from the Mn *d* states and occupied minority $t_{2g}$ almost no minority Mn *d* states. This is consistent well with the conclusion that the lower $t_{2g}$ orbitals are dominated by O *p* orbitals and the higher by Mn *d* orbitals. This is also consistent well with the traditional Hund rule.

There are many factors affecting the spin injection efficiency in spintronics-semiconductor paradigm, such as the lattice mismatch, interfacial scattering, symmetry breaking, formation of alloys and so on. Theoretical and practical studies show that the spin injection efficiency is low when directly inject spin-polarized electrons into semiconductor, even for diluted magnetic semiconductors and half-metal. New evidences demonstrate that ballistic hot-electron injection would be a good method, where the spin injection efficiency and quality is determined by the spintronics thin films served as spin filter [1]. For the hot-electrons are not always coherent, a wide range of energy with HMF would be necessary for good quality spin injection. Interestingly, Mn doped $ZrO_2$ show HMF in large range energy. Moreover, delta-doping (including other heterogeneous doping profiles) would expand HMF to higher energy than homogeneous doping owing to Jahn-Teller effect. A point should be mentioned, the spin direction would undergo a up-zero-down transition via modulating the hot-electrons energy from lower to higher, this is the typical character of magnetic semiconductors. We think which would be very useful and maybe find use in spin direction controllment.

**4. Conclusions**

In summary, using first-principles calculations we investigate the electronic structure and magnetism in 25% Mn doped cubic $ZrO_2$ with several homogeneous and heterogeneous doping profiles. The results indicate that all studied doping profiles show half-metallic ferromagnetism (HMF), and delta-doping is energy favorable while homogeneous doping is most ferromagnetic stable. Using crystal field theory, we explain the formation of HMF and the Jahn-Teller effect on DOS of studied doping profiles. The delta-doping shows larger energy rang of HMF than other doping profiles. Finally, we speculate the potential use in spin direction controllment for Mn doped $ZrO_2$.

**Acknowledgments**

The work is partially supported by the Postgraduate Innovation Foundations of China University of Petroleum (No. B2008-8), and Xingtao Jia acknowledges the suggestion from Professor A. Ernst at Max-Planck-Institut fur Mikrostrukturphysik and T. Archer at Trinity Coolege.

# References

[1] Appelbaum I, Huang B and Monsma D J 2007 *Nature* **447** 295
[2] Huang B, Monsma D J and Appelbaum I 2007 *Appl. Phys. Lett.* **91** 072501
[3] Huang B, Monsma D J and Appelbaum 2007 *Phys. Rev. Lett.* **99** 177209
[4] Jansen R 2007 *Nat. Phys.* **3** 521
[5] Schmehl A, Vaithyanathan V, Herrnberger A, Thiel S, Richter C, Liberati M, Heeg T, Röckerath M, Kourkoutis L F, Mühlbauer S, Böni P, Muller D A, Barash Y, Schubert J, Idzerda Y, Mannhart J, Schlom D G. 2007 *Nat Mater.* **6** 798-9.
[6] Felser C, Fecher G H and Balke B 2007 Angew. Chem. Int. Ed. 46 668
[7] Ostanin S, Ernst A, Sandratskii L M, Bruno P, Däne M, Hughes I D, Staunton J B, Hergert W, Mertig I and Kudrnovský 2007 *Phys. Rev. Lett.* **98** 016101
[8] Archer T, Pemmaraju C Das and Sanvito S 2007 *J. Magn. Magn. Mater.* **316** e188
[9] Shi H and Duan Y 2008 *J. Appl. Phys.* **103** 073903
[10] Ramachandra Rao M S, Dhar S, Welz S J, Ogale S B, Kundaliya Darshan C, Shinde S R, Lofland S E, Metting C J, Erni R, Browning N D and Venkatesan T 2004 eprint arXiv:cond-mat/0405378
[11] Venkatesan M, Fitzgerald C B and Coey J M D 2004 *Nature* **430** 630
[12] Ramachandra Rao M S, Kundaliya D C, Ogale S B, Fu L F, Welz S J, Browning N D, Zaitsev V, Varughese B, Cardoso C A, Curtin A, Dhar S, Shinde S R, Venkatesan T, Lofland S E and Schwarz S A 2006 *Appl. Phys. Lett.* **88** 142505
[13] Bouzerar G and Ziman T 2006 *Phys. Rev. Lett.* **96** 207602
[14] Herwadkar A, Lambrecht W R L and Schilfgaarde M van 2008 *Phys. Rev. B* **77** 134433
[15] Miura Y, Shirai M and Nagao K 2004 *J. Phys.: Condens. Matter* **16** s5735
[16] Liu Q, Yan W, Wei H, Sun Z, Pan Z, Soldatov A V, Mai C, Pei C, Zhang X, Jiang Y, and Wei S, *Phys. Rev. B* **77** 245211
[17] Kaspar T C, Droubay T, Heald S M, Engelhard M H, Nachimuthu P and Chambers S A 2008 *Phys. Rev. B* **77**, 201303
[18] Collins B A, Chu Y S, He L, Zhong Y and Tsui F 2008 *Phys. Rev. B* **77** 193301
[19] Schubert E F, Proetto C R, Ploog K H, Makimoto T, Horikoshi Y, Ritter D, Eisele I, Gossmann H -J, Luftmann H S, Gossmann H -J, Newman R C, Koenraad P M, Yao H, Wagne J, Richards D, Masselink W T, Koenraad P M, Asche M, Headrick R L, Feldman L C, Weir B E, Hong W -P, Nakagawa K, Yamaguchi K and Malik R J 1995 *Delta-doping of Semiconductors* ed E F Schubert (Cambridge: Cambridge University Press) pp 1-601
[20] Goniakowski J, Finocchi F and Noguera C 2008 *Rep. Prog. Phys.* **71** 016501
[21] Yanagisawa K, Takeuchi S, Yoshitake H, Onomitsu K and Horikoshi Y 2007 *J. Cryst. Growth* **301-302**, 634
[22] Nazmul A M, Amemiya T, Shuto Y, Sugahara S and Tanaka M 2005 *Phys. Rev. Lett.* **95** 017201
[23] Jia X, Yang W, Li H and Qin M 2008 *J. Phys. D: Appl. Phys.* **41** 115004
[24] Wang H –Y and Qian M C 2006 *J. Appl. Phys.* **99** 08D705
[25] Qian M C, Fong C Y, Liu K, Pickett W E, Pask J E and Yang L H 2006 *Phys. Rev. Lett.* **96** 027211
[26] Zhu L Q, Fang Q, He G, Liu M and Zhang L D 2006 *J. Phys. D: Appl. Phys.* **39** 5285
[27] Zhu J, Li Y R and Liu Z G 2004 *J. Phys. D: Appl. Phys.* **37** 2896
[28] Peacock P W and Robertson J 2004 *Phys. Rev. Lett.* **92** 057601
[29] Wang S J, Huan A C H, Foo Y L, Chai J W, Pan J S, Li Q, Dong Y F, Feng Y P and Ong C K 2004 *Appl. Phys. Lett.* **85** 4418.
[30] Okabayashi J, Toyoda S, Kumigashira H, Oshima M, Usuda K, Niwa M and Liu G L 2004 *Appl. Phys. Lett.* **85** 5959.
[31] Navrotsky A 2005 *J. Mater. Chem.* **15** 1883.
[32] Segall M D, Lindan P L D, Probert M J, Pickard C J, Hasnip P J, Payne S M C and Clark J 2002 *J. Phys.:Cond. Matt.* **14** 2717
[33] Vanderbilt D 1990 *Phys. Rev. B* **41** 7892
[34] Perdew J P, Burke K and Ernzerhof M 1996 *Phys. Rev. Lett.* **77** 3865
[35] Hordequin C, Ristoiu D, Ranno L and Pierre J 2000 *Eur. Phys. J. B* **16** 287

**Table 1.** Calculated relevant properties in full relaxed 25% Mn doped $ZrO_2$ with different doping profiles: total magnetic moment ($m_{tot}$), total energy (with respect to N112), and ferromagnetic stabilization energy (energy difference between antiferromagnetic and ferromagnetic states, $\Delta E_{FA}=E_{FM}-E_{AFM}$).

|      | $d$ (Å) | $m_{tot}$ ($\mu_B$) | $E_{FM}$ (meV) | $\Delta E_{FA}$ (meV) |
|------|---------|---------------------|----------------|------------------------|
| N112 | 5.061   | 3.0                 | 0              | 34.4                   |
| N002 | 5.059   | 3.0                 | -91.6          | 173.1                  |
| N011 | 3.497   | 3.0                 | -170.7         | 121.2                  |
| N110 | 3.574   | 3.0                 | -273.7         | 137.9                  |

**Figure captions**

**Figure 1.** Doping profiles of 25% Mn doped $ZrO_2$. Therein, one Mn atom sites at the origin and another at the nearest positions along different directions.

**Figure 2.** Calculated total Density of States (DOS) of 25% Mn doped $ZrO_2$ with different doping profiles.

**Figure 3.** Scheme of the *p-d* hybridization in Mn doped $ZrO_2$. Therein, $d_1…d_5$ denote $d_x$, $d_y$, $d_z$, $d_{x^2-y^2}$ and $d_{z^2}$ respectively; and $p_1$, $p_2$, $p_2$ denote $p_x$, $p_y$, and $p_z$ respectively.

**Figure 4.** Calculated *total* and partial and element resolved Density of States (DOS) of 25% Mn delta-doped $ZrO_2$.

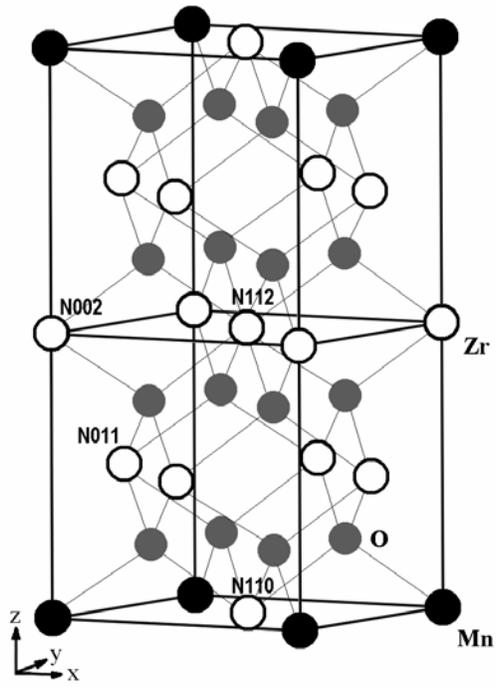

**Figure 1** X. Jia et al.

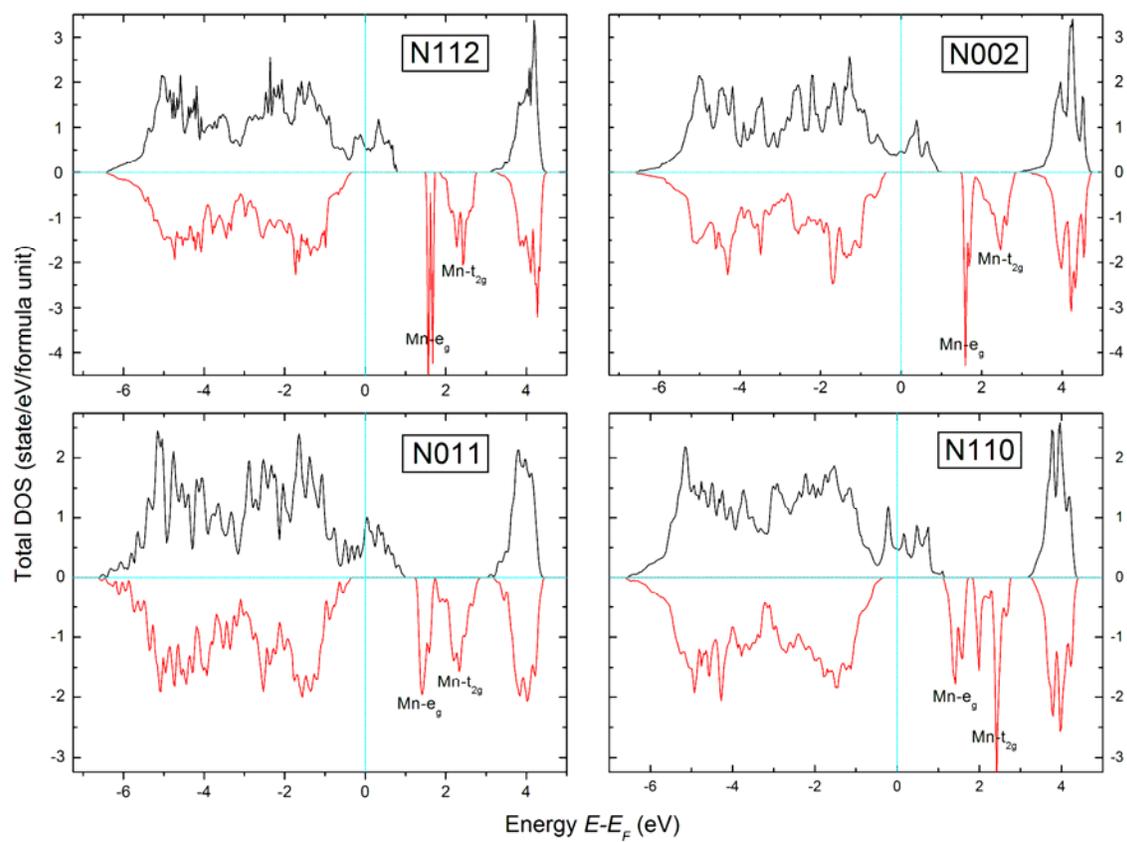

**Figure 2** X. Jia et al.

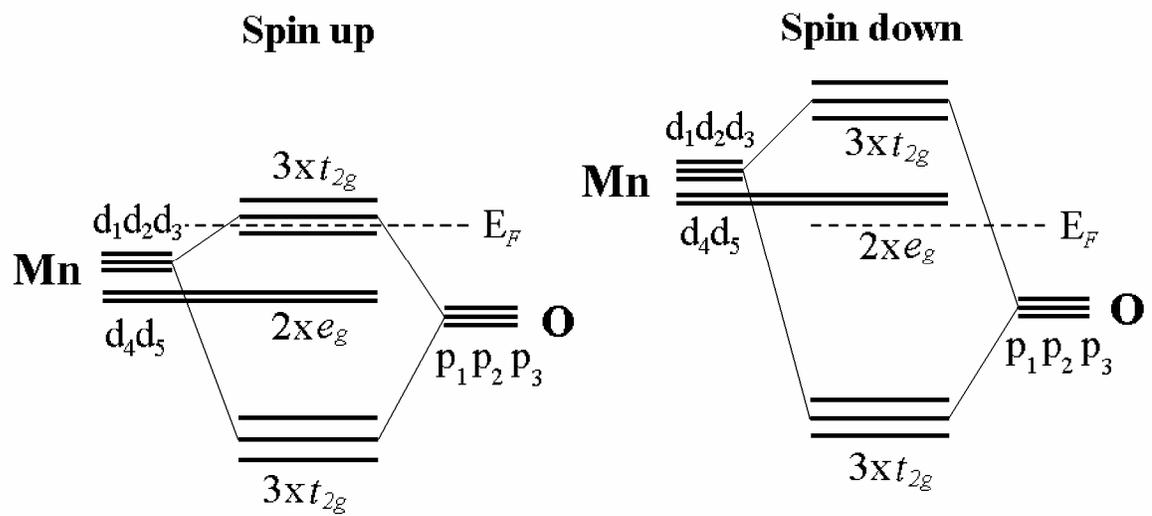

**Figure 3** X. Jia et al.

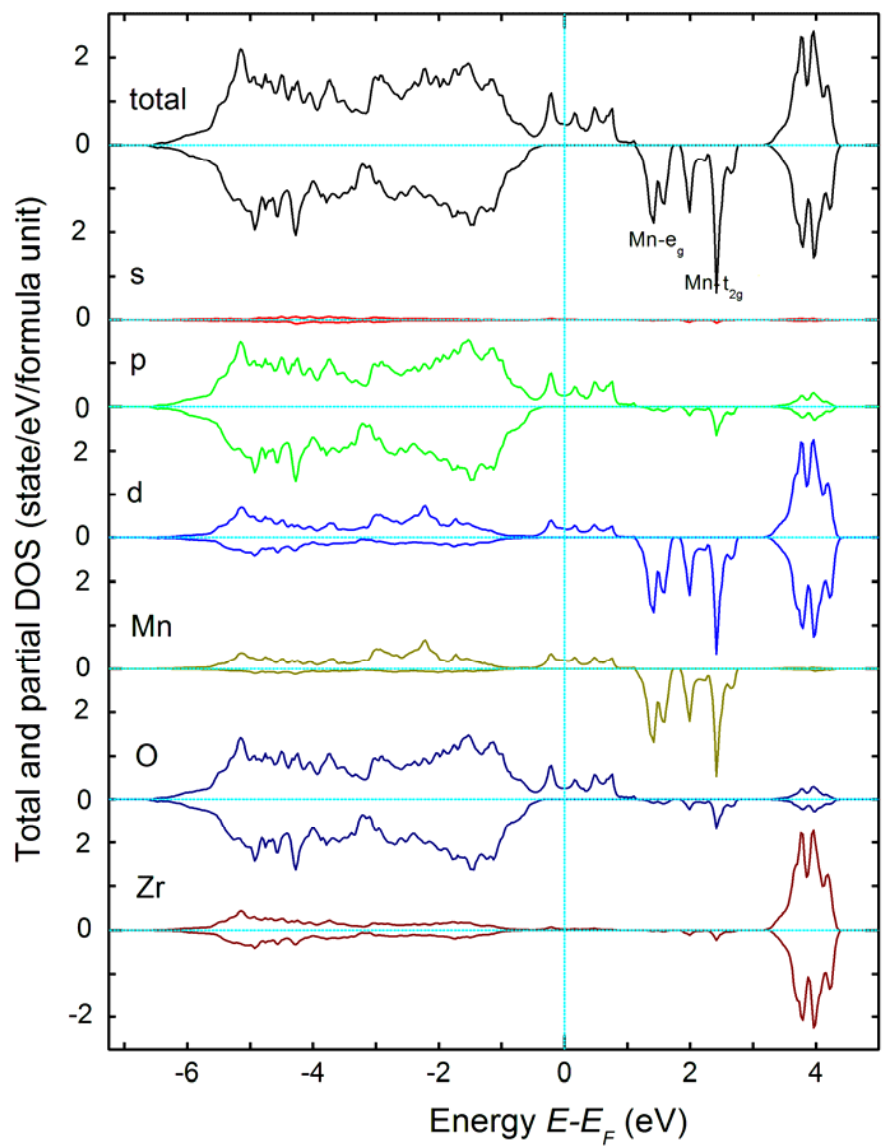

**Figure 4** X. Jia et al.